\documentclass{sig-alternate}
\usepackage{mathptmx} 

\usepackage{comment}
\usepackage{fancyhdr}
\usepackage[normalem]{ulem}
\usepackage[hyphens]{url}
\usepackage[sort,nocompress]{cite}
\usepackage[final]{microtype}
\usepackage[keeplastbox]{flushend}
\usepackage[bookmarks=true,breaklinks=true,letterpaper=true,colorlinks,citecolor=blue,linkcolor=blue,urlcolor=blue]{hyperref}
\usepackage{siunitx}

\usepackage[braket, qm]{qcircuit}
\usepackage{graphicx}
\usepackage{subfig}
\usepackage{float}
\usepackage{multicol}
\usepackage{multirow}
\usepackage{units}
\usepackage{xcolor, soul}
\newcommand{\colorhl}[2][yellow]{#2}

\usepackage{ifthen}
\usepackage{amssymb}
\usepackage{xcolor}

\newboolean{showcomments}
\setboolean{showcomments}{true}

\makeatletter
\newcommand{\mynote}[3]{%
  \ifthenelse{\boolean{showcomments}}{%
   \fbox{\bfseries\sffamily\scriptsize#1}%
   {\small$\blacktriangleright$\textsf{\emph{\color{#3}{#2}}}$\blacktriangleleft$}}%
  {%
   \@bsphack
   \@esphack
  }%
}
\makeatother

\definecolor{asparagus}{rgb}{0.53, 0.66, 0.42}

\newcommand{\test}[1][]{%
\text{\texttt{iSWAP}}%
\ifthenelse{\equal{#1}{}}{}{^{#1}}%
}

\newcommand{\sqisw}[1][]{\sqrt[#1]{\text{\texttt{iSWAP}}}}
\newcommand{\isw}[1][]{\text{\texttt{iSWAP}}^{#1}}
\newcommand{\sw}[1][]{\text{\texttt{SWAP}}^{#1}}

\newcommand{\cnot}{\texttt{CNOT}}
\newcommand{\cx}{\texttt{CX}}

\newcommand{\fsim}{\texttt{FSIM}}
\newcommand{\syc}{\texttt{SYC}}
\newcommand{\CR}{\texttt{CR}}
\newcommand{\isfam}{$\sqrt[n]{\text{\texttt{iSWAP}}}$ }

\newcounter{obsctr}
\newenvironment{observation}[0]{\refstepcounter{obsctr}\par\medskip
\noindent\textbf{Observation~\theobsctr.} \rmfamily}{\medskip}

\title{Co-Designed Architectures for Modular Superconducting Quantum Computers\vspace{-0.5in}} 
\author{Evan McKinney$^\dagger$, Mingkang Xia$^\ddag$, Chao Zhou$^\ddag$, Pinlei Lu$^\ddag$, Michael Hatridge$^\ddag$, Alex K. Jones$^\dagger$\\
\normalsize \{evm33,mix20,chz78,pil9,hatridge,akjones\}@pitt.edu\\
\normalsize $^\dagger$Department of Electrical and Computer Engineering, $^\ddag$Department of Physics and Astronomy\\
\normalsize University of Pittsburgh}
\begin{document}
\maketitle
\pagestyle{plain}



\begin{abstract}

Noisy, Intermediate Scale Quantum (NISQ) computers have reached the point where they can show the potential for quantum advantage over classical computing. Unfortunately, NISQ machines introduce sufficient noise that even for moderate size quantum circuits the results can be unreliable. We propose a collaboratively designed superconducting quantum computer using a Superconducting Nonlinear Asymmetric Inductive eLement (SNAIL) modulator. The SNAIL modulator is designed by considering both the ideal fundamental qubit gate operation while maximizing the qubit coupling capabilities. 
\colorhl[yellow]{First, the SNAIL natively implements $\sqrt[n]{\text{\texttt{iSWAP}}}$ gates 
realized through proportionally scaled pulse lengths.  This naturally includes $\sqrt{\text{\texttt{iSWAP}}}$, which provides an advantage over $\texttt{CNOT}$ as a basis gate.  Second, the SNAIL enables high-degree couplings that allow rich and highly parallel qubit connection topologies without suffering from frequency crowding.  Building on our previously demonstrated SNAIL-based quantum state router we propose a quantum 4-ary tree and a hypercube inspired corral built from interconnected quantum modules.  We compare their advantage in data movement based on necessary \texttt{SWAP} gates to the traditional lattice and heavy-hex lattice used in latest commercial quantum computers.  
We demonstrate the co-design advantage of our SNAIL-based machine with $\sqrt{\text{\texttt{iSWAP}}}$ basis gates and rich topologies against $\texttt{CNOT}$/heavy-hex and $\texttt{FSIM}$/lattice for 16-20 qubit and extrapolated designs circa 80 qubit architectures.  We compare total circuit time and total gate count to understand fidelity for systems dominated by decoherence and control imperfections, respectively.  Finally, we provide a gate duration sensitivity study on further decreasing the SNAIL pulse length to realize $\sqrt[n]{\text{\texttt{iSWAP}}}$ qubit systems to reduce decoherence times.}
\end{abstract}


\section{Introduction}
Quantum Computers (QCs) leverage the physics of quantum information with the promise to change the computing landscape by solving problems that that are intractable for classical computers.  The ingenuity of QCs comes from quantum superposition and entanglement which, unlike classical computers, allows the QC core computing element, or \textit{qubit}, to conceptually interact with all other qubits simultaneously.  
However, to achieve practical quantum-advantage requires fault-tolerance, \textit{i.e.,} passing a threshold of sufficient QC size and fidelity rates to build error-correcting schemes 
~\cite{gottesman1997stabilizer}. The field is currently in the the Noisy Intermediate-Scale Quantum (NISQ) era where quantum machines with more than a hundred qubits exist and the fidelities of quantum operations among one or more qubits, typically referred to as \textit{gates}, above $99.9\%$ are possible,
yet remain too small and sensitive to error to perform error-correction
~\cite{preskill2021quantum}. 
Therefore, NISQ machines are crucially constrained by the duration of the circuits to limit gate noise and qubit decoherence. 



NISQ QCs operate by qubit coupling mechanisms which produce different gate operations and neighborhoods of qubit connectivity. Practically, this qubit-qubit coupling arises from a physical connection between them, such as a simple capacitive coupling to a more elaborate nonlinear circuit.  We will refer to them generally as \textit{modulators}, which may target pairs of qubits via layout geometry, unique frequencies or frequency differences, and which, together with applied control signals, govern the fundamental gate operations implemented in the quantum computer. Due to the strict constraints of duration and decoherence, it is necessary to advance the design of modulators to produce high-fidelity qubits and couplings.

\begin{figure}[h]
    \centering
    \includegraphics[width=\columnwidth]{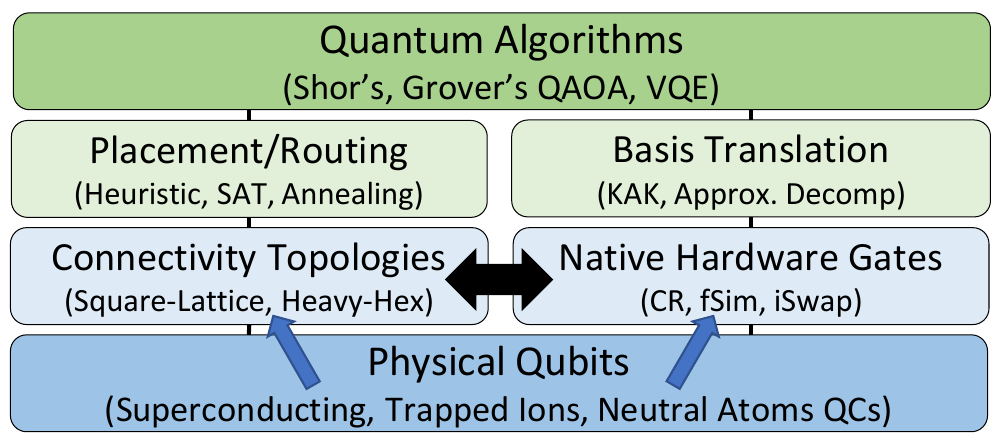}
    \caption{\colorhl[yellow]{Traditional quantum computing stack.  The SNAIL allows co-design of rich connectivity topologies and native hardware \textit{basis} gates for improved basis translation, placement, and routing of important quantum algorithms.}}
    \label{fig:co-design}
\end{figure}

\colorhl[yellow]{The realized potential of quantum computing is furthered when elements of the quantum computing stack, shown in Fig.~}\ref{fig:co-design}, are designed synergistically.  This requires development of the physical mechanisms and modulators, which realize the \textit{physical qubits}, and the \textit{native hardware gates} and \textit{connectivity topologies}, respectively are collaboratively designed with the \textit{quantum algorithms} we want to implement.  Thus, these quantum circuits more successfully execute a quantum program with higher reliability to advance the applicability of NISQ computers.

In this paper we propose \textit{\textbf{a novel co-designed superconducting QC architecture}} based on a ``SNAIL'' (Superconducting Nonlinear Asymmetric Inductive eLement) quantum modulator~\cite{frattini20173}.  We leverage our recent experimental demonstration of a SNAIL-based quantum state router~\cite{zhou2021modular} and prototype SNAIL-based four qubit modules to characterize and construct modular quantum topologies such as various designs based on a 4-ary tree with native $\isw$-family gates. \colorhl[yellow]{The SNAIL modular QC technology is in its early stages of its long-term potential. Our co-design methodology is important for scaling considerations ahead of what has been physically constructed, to help discover the fundamental advantages in comparison to prior art while perfecting the physical components and modules at a small scale}~\cite{Quantum-co-design}.

We explore the benefit of $\sqisw$ 
as a basis operation as an improved decomposition alternative to \cnot{}, or $\syc$ gates~\cite{huang2021towards}.  
We then explore how decomposition to $\sqisw[n]$ 
where $n\geq3$ can further improve fidelity through reduced pulse time. Furthermore, we consider alternate topologies in a search for designs which are both physically reasonable with demonstrated technology and maximize the computational power of SNAIL-based modular NISQ machines.  All results are benchmarked against existing large-scale machines from Google~\cite{Arute2019} based on a Square-Lattice of qubit couplings and the $\fsim$ gate family and IBM's recent ``Heavy-Hex" based machines with native $\CR$ gates~\cite{chamberland2020topological}.



In particular, our contributions are as follows:
\begin{itemize}
\vspace{-.05in}
    \item Demonstrate novel and scalable high connectivity 4-ary tree and Corral topology modules enabled through the co-designed QC using SNAIL modulators.
\vspace{-.05in}
    \item Explore $\sqisw[n]$ 
    as the basis gate in our co-designed QC to increase fidelity through improving overall time to algorithm completion.
\vspace{-.05in}
    \item Demonstrate our SNAIL-based modular QCs potential improvement for moderate 16-20 qubit QCs versus normalized versions of IBM and Google architectures as representative systems on a set of NISQ-algorithms.
\vspace{-.05in}
    \item Explore the scalability of collaboratively designed modular superconducting QCs for large numbers of qubits.
\end{itemize}

We find that on an average of Quantum Volume (QV) circuits ranging from 16 to 80 qubits, a hypercube topology with a $\sqisw$ basis gate requires $3.16\times$ less total 2Q gates and $6.11\times$ less total duration-dependent 2Q gates than a Heavy-Hex topology with a $\cnot$ basis. Additionally, we find that for randomly sampled 2Q unitaries and for a $99\%$ fidelity $\isw$ basis, $\sqisw[4]$ decreases infidelity on average by $25\%$ compared to $\sqisw$.




\section{Background and Preliminaries}
In this section we provide some relevant background on quantum computing principles and relevant preliminary statements about co-design methodologies. 
\subsection {Transmon Qubits}
\label{sec:transmon}
Qubits have been realized from a variety of quantum systems, including atoms, electrons, nuclear spins, etc. In this work we focus on superconducting qubits which are nano-fabricated, nonlinear microwave-frequency circuits \cite{devoret2013superconducting}. The nonlinear element of choice in these circuits is the Josephson junction (JJ), which can be thought of as a non-linear inductor.  The most commonly used superconducting qubit is the transmon, which consists simply of a JJ shunted with a large capacitance\cite{koch2007charge}.  The transmon is ubiquitous because it is insensitive to charge and flux noise and can achieve very high coherence~\cite{place2021new, wang2022towards}.  It is also easy to fabricate, control, and couple to other circuit elements\cite{blais2021circuit}.  

Superconducting qubits are almost always controlled and read out by embedding them in a linear oscillator; this platform of qubit + cavity as the basic unit goes by the name circuit Quantum Electrodynamics or cQED\cite{blais2021circuit}.  

The cavity 
is coupled to the qubit dispersively via the 
``cross-Kerr term'' 
which shifts the mode of the cavity 
when the qubit gains a photon.  
This allows for a pulse transiting the cavity to encode the state of the qubit, and is the basis for qubit readout.  More, dispersive interactions and cross-Kerr interactions can also be used to couple qubits to each other, to a resonator, or to another nonlinear object, and so are often the physical basis for two qubit gates as well.

\subsection{Quantum Gates}
Quantum gates are unitary matrix operations that act on quantum state vectors.  
In general, for NISQ machines single-qubit gates (1Q) and two-qubit gates (2Q) form the building blocks of quantum circuits \cite{nielsen2002quantum}. A native QC gate set, analogous to a classical computer's instruction set architecture, defines which unitary operations are available to use on a particular machine. In the same way that \texttt{NAND} and \texttt{NOR} gates are universal digital logic gates, as either gate can implement any boolean logic expression, the requirement for a quantum gate set to be universal is that it can implement any arbitrary unitary to arbitrary accuracy. A basic universal gate set consists of arbitrary single qubit gates plus a single two qubit gate, most often $\cnot$\cite{barenco1995universal}, shown in Eq.~\ref{cnot}.  




The $\cnot$ gate acts like a classical reversible-$\texttt{XOR}$ gate, with the crucial difference that its inputs can be in superposition.
Like many other universal 2Q gates, $\cnot$ is a perfect entangler\cite{PhysRevA.70.052313}, meaning that it can take a separable state
to a maximally entangled state.
Entangled states are ones in which we can only describe the state of two qubits jointly and losing information about one qubit destroys information in its entangled partner.  A different, potentially more convenient, physical coupling of a QC may result in a different \textit{basis} gate. For example the family of fractional-$\isw$ gates, as in Eq.~\ref{iswap}, for which $\isw$ and $\sqisw$ are also universal gates. 



\vspace{-.1in}
\begin{equation}
    \label{cnot}
    \text{\cnot} =
    \begin{bmatrix}
    1 & 0 & 0 & 0\\
    0 & 1 & 0 & 0\\
    0 & 0 & 0 & 1\\
    0 & 0 & 1 & 0
    \end{bmatrix}
\end{equation}

\begin{equation}
    \label{iswap}
    \sqisw[n] =
    \begin{bmatrix}
    1 & 0 & 0 & 0\\
    0 & \cos({\pi/2n}) & \mathit{i}\sin({\pi/2n}) & 0\\
    0 & \mathit{i}\sin({\pi/2n}) & \cos({\pi/2n}) & 0\\
    0 & 0 & 0 & 1
    \end{bmatrix}
\end{equation}



\sloppy 

\subsection{Gate Translation and Decomposition}
\label{section:decomp}

Translating a quantum program between basis gates such as $\cnot$ or $\sqisw$ uses a process called gate decomposition.  
Decomposition converts arbitrary unitaries from the quantum program into an equivalent sequence of unitaries only compromised of gates from the basis set of the target system. Note, in the superconducting quantum computers, 1Q gates are generally much faster and higher fidelity than 2Q gates and are often treated as negligible or perfect~\cite{vzgates}. 

Prior work has shown at most only 3 $\sqisw$ gates plus four interleaved rounds of 1Q gates (given as $U_1$ to $U_8$) are required to decompose any 2Q unitary~\cite{huang2021towards}, depicted in Eq.~\ref{sqiswap-decomp}. 
 \colorhl{Cartan's KAK is an efficient exact decomposition method that uses geometric relationships between 2Q unitaries to solve for the interleaved 1Q gate values}, where intuitively unitary transformations are thought of as movement inside a periodic coordinate basis known as the Weyl Chamber~\cite{tucci2005introduction}.  KAK decomposition shows that many unitaries may in fact be decomposed into only 2 $\sqisw$.  As similar decomposition is more widely known for $\cnot$, while a higher portion of unitaries require 3 $\cnot$s than 3 $\sqisw$s~\cite{huang2021towards}.

\begin{equation}
\label{sqiswap-decomp}
\resizebox{\columnwidth}{!}{
    \Qcircuit @C=0.4em @R=0.4em @!R {
	 	&\multigate{2}{\mathrm{U}} &\qw & & & \gate{\mathrm{U_1}} & \multigate{2}{\mathrm{\sqisw}} & \gate{\mathrm{U_3}} &
	 	\multigate{2}{\mathrm{\sqisw}} & \gate{\mathrm{U_5}} &
	 	\multigate{2}{\mathrm{\sqisw}} & \gate{\mathrm{U_7}} & \qw
	 	\\
	 	& & \push{\rule{0em}{0em}=\rule{0em}{0em}}
	 	\\
	 	&\ghost{\mathrm{U}} &\qw & & & \gate{\mathrm{U_2}} & \ghost{\mathrm{\sqisw}} & \gate{\mathrm{U_4}} &
	 	\ghost{\mathrm{\sqisw}} & \gate{\mathrm{U_6}} &
	 	\ghost{\mathrm{\sqisw}} & \gate{\mathrm{U_8}} &\qw
	 	\\
	 }}
\end{equation}

While it is possible to convert circuits in a $\cnot$ basis into the $\sqisw$ basis, it is more efficient to decompose the original unitaries of the algorithm directly into the basis gates.  \colorhl[yellow]{$\cnot$ can be realized by 2 $\sqisw$.  In the best case a unitary can be realized in 2 $\cnot$.  Conversion would require 4 $\sqisw$, while direct decomposition would require, at worst, 3 $\sqisw$.  This is particularly bad for unitaries that require 3 $\cnot$ that can be directly decomposed to 2 $\sqisw$, for which conversion would require 6 $\sqisw$, making conversion a 1.3--3$\times$ increase in 2Q gates over direct decomposition.} 
When direct decomposition for a basis gate has not been solved, an alternative is to use approximate decomposition using numerical optimization techniques. 
An optimizer is run to minimize the distance, or variance between the requested 2Q gate and the approximation, on the candidate circuit structure, either by converging on a set of 1Q gate parameters or expanding the circuit template to include more target 2Q gate instances~\cite{smith2021leap}. 
%
%
The primary limitation of numerical decomposition is that for quantum circuits with moderate to large gate counts, it can take orders of magnitude longer time compared to the closed-form substitution rules. 
Nonetheless, numerical decomposition can be helpful to study new basis gates without needing to mathematically derive analytic decomposition rules for each new 2Q basis gate, such as for the fractional durations \isfam gates.

In the next section the coupling neighborhoods, formed from the combination of modulator and physical connections, are described in more detail.  

\subsection{Qubit Coupling and Topologies}
To perform both 1Q and 2Q gates, the control signals applied to all qubits and their associated modulators must be unique.  
For example, in the case of two qubits coupled via a central modulator, a 1Q gate on qubit 1 must operate without creating spurious 2Q gates or drive qubit 2. This is typically accomplished via a combination of (1) spatial selectivity, in which drive lines couple only to a single qubit or modulator and/or (2) frequency selectivity, in which, for example nearest-neighbor qubits have deliberately spaced frequencies to suppress cross-talk among 1Q gates. 

Conceptually there is no limit to the density of qubit coupling connections in the neighborhood.  In practice, connectivity is limited by these requirements and is typically relatively small, with qubits having 2-4  neighbors and 2-6 couplings. 

Accordingly, a graph $G = \{V,E\}$ is used to represent the organization of the quantum computer where physical qubits form the vertices in $V$ and a coupling capability to perform 2Q gates between qubits are edges in $E$.

A complicating factor in discussing QC topology is that \emph{the choice of gate type and coupling topology are not independent}, as they are both determined by the choice of modulator.  Thus, in the remainder of this section we will introduce the most common qubit modulators as comparison points. 

\subsubsection{Cross-resonance gate}

The first modulator/gate 
originally proposed by IBM \cite{chow2011simple}, is the `cross-resonance' ($\CR$) or \texttt{ZX} gate. It 
utilizes the dispersive cross-Kerr interaction~\cite{blais2021circuit} between two qubits
to realize a 2Q 
gate.  $\CR$ drives the second qubit at the first qubit's frequency. In the driven frame, the interaction this drive creates couples the Z-component of the first qubit to an X rotation on the second qubit ($H= Z_1 X_2$). This action resembles a $\cnot$ gate 
but 
is most often translated into a true $\cnot$ by adding 1Q gates as noted in Eq.~\ref{CR:circuit}. In quantum computers the linkage between the two qubits is affected via a third mode~\cite{chow2011simple, kandala2019error, chow2021ibm}. 

\begin{equation}
    \label{equation:CR}
\hspace{-.1in}
    \texttt{ZX}(\theta) = \begin{bmatrix}
    \cos{\theta/2} & 0 & -\mathit{i}\sin{\theta/2} & 0\\
    0 & \cos{\theta/2} & 0 & \mathit{i}\sin{\theta/2}\\
    -\mathit{i}\sin{\theta/2} & 0 & \cos{\theta/2} & 0\\
    0 & \mathit{i}\sin{\theta/2} & 0 & \cos{\theta/2}
    \end{bmatrix}
\end{equation}

This gate has successfully realized high fidelities in large systems, and is used throughout IBM's fleet of QCs.  The challenges this gate faces are that: (a) the un-driven cross-Kerr interaction creates $Z_1 Z_2 $ errors continuously while off, (b) the qubits should be close in frequency, which does not allow for many-to-many interactions, and (c) given the former, $\CR$ gates require very precise fabrication to avoid cross-talk, which has motivated IBM's shift to more sparsely connected Heavy-Hex architectures~\cite{chow2021ibm}.

\begin{equation}\label{CR:circuit}
\Qcircuit @C=1.0em @R=0.2em {
& \ctrl{2} & \qw & & & \multigate{2}{\mathrm{ZX}\,(\mathrm{\frac{\pi}{2}})} & \gate{\mathrm{S^\dagger}} & \qw\\
&&& \push{\rule{0em}{0em}=\rule{0em}{0em}} & & {}\\
& \targ & \qw & & & \ghost{\mathrm{ZX}\,(\mathrm{\frac{\pi}{2}})} & \gate{\mathrm{\sqrt{X}^\dagger}} & \qw\\
}
\end{equation}


\subsubsection{Direct photon exchange: \texorpdfstring{$\isw$}{iSwap} and \texorpdfstring{$\fsim$}{fSim} gates}
\label{section:photon-exchange}

Another category of gate is direct photon exchange. In this two qubits are coupled resonantly for a period of time to exchange light via the photon-exchange interaction. 
To form a gate requires turning this interaction on and off.  Typically either the qubit frequencies must be tuned together in frequency to exchange, and then far apart to stop the interaction~\cite{dicarlo2009demonstration, barends2014superconducting}, or by using a `tunable coupler' in between~\cite{bialczak2011fast}. Direct exchange naturally yields $\isw$-family gates.  The exponent of the unitary is determined by the combination of interaction strength and gate duration.
The coupler approach has been adopted by Google Quantum AI~\cite{Arute2019} in their Sycamore ($\syc$) architecture among other groups~\cite{sung2021realization,li2020tunable, niskanen2007quantum,blais2003tunable}.  $\syc$ gates accrue a phase on the $\ket{11}$ state in addition to $\isw$, termed $\fsim$ given in Eq.~\ref{fSim}.  $\theta$ and $\phi$ are determined by the pulses applied to the coupler.
\begin{equation}
    \label{fSim}
    \fsim(\theta, \phi) = \begin{bmatrix}
    1 & 0 & 0 & 0\\
    0 & \cos{\theta} & -\mathit{i}\sin{\theta} & 0\\
    0 & -\mathit{i}\sin{\theta} & \cos{\theta} & 0\\
    0 & 0 & 0 & e^{-\mathit{i}\phi}
    \end{bmatrix}
\end{equation}
$\syc$ sets $\theta=\pi/2, \phi=\pi/6$. The $\fsim$ gate set yields respectable on/off ratios, but suffers from challenges due to: (a) the difficulty of implementing rapid, extremely precise base-band flux control~\cite{Kafri2017} and strong sensitivity to flux noise in these controls\cite{mccourt2022learning}, (b) the requirement for equal-frequency qubit and concomitant qubit-qubit cross talk issues, and (c) the recent demonstration of strong flux-noise based noise and qubit dephasing in the couplers~\cite{yan2018tunable}.  $\sqisw$ is realized by setting $\theta=-\pi/4, \phi=0$.



\subsubsection{Data Movement on the Topology}

Movement between physical qubits is accomplished using non-entangling $\sw$ gates.  $\sw$s can be directly decomposed to 3 $\cnot$s, as well as 3 $\sqisw$, although the latter requires additional 1Q unitaries. 
A quantum algorithm, represented as a graph $G' = \{V',E'\}$, is mapped to hardware topology by embedding $G'$ in $G$ and inducing $\sw$ gates when edges cannot be directly realized. As $\sw$ gates consist themselves of noisy 2Q hardware gates, it is important to reduce this cost to maximize the overall fidelity of the circuit. Increasing the connectivity in the QC topology will reduce the overhead $\sw$ gate cost compared to a sparsely connected graph, similarly impacting fidelity. 

\subsubsection{Common Qubit Topologies}
\label{section:common-topologies}
\begin{figure}[h]
    \centering
    \includegraphics[width=\columnwidth]{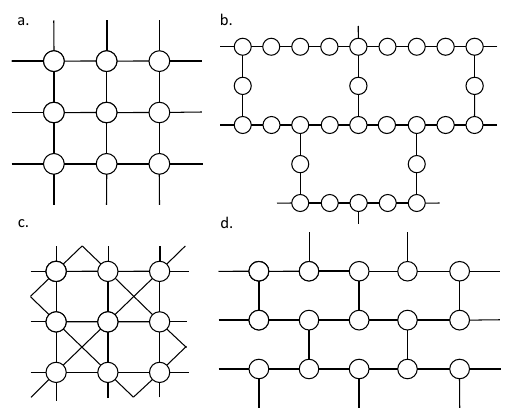}
    \vspace{-.10in}
    \caption{Standard Qubit Coupling Topologies including (a). Square-Lattice, (b). Heavy-Hex, (c). Lattice with Alternating Diagonals, (d). Hex-Lattice}
    \label{fig:generic_topology}
\end{figure}

A simple topology that couples qubits to each of their four nearest-neighbors is the Square-Lattice, Fig.~\ref{fig:generic_topology}a, which is regular and straight-forward to expand. 
IBM's early Penguin machines attempted higher connectivities with diagonals on alternating tiles of the Square-Lattice, Fig.~\ref{fig:generic_topology}c, with limited success due to issues of frequency crowding at the cost of fidelity. For this reason, IBM has over time reduced the connectivity, \colorhl[yellow]{moving to a Hex-Lattice}, Fig.~\ref{fig:generic_topology}d, and now currently to Heavy-Hex topologies, Fig.~\ref{fig:generic_topology}b~\cite{nation_paik_cross_nazario_2021}. All of these topologies have been demonstrated experimentally with varying degrees of success using $\fsim$ and/or $\CR$ drive protocols.

These topologies are guided by the constraints of two-dimensional circuits whose modulators do not, in general cross each other or span long distances across the chip, despite however beneficial that might be.  
Instead, a richly connected topology of particular interest in applications of networking and parallel computing is the hypercube, or n-dimension cube. Hypercubes are of particular theoretic interest to qubit coupling topologies because for $2^n$ nodes, both the number of edges incident on every node and the distance between any pair of nodes are exactly $n$, hence scaling regularly and efficiently the neighborhoods of qubit couplings and induced $\sw$ operations. Implementing such a topology requires us to be able to link a given qubit to $n$ neighbors, which in turn requires a modulator with with this connectivity.  


\begin{figure}[tbp]
\vspace{-.15in}
    \centering
    \includegraphics[width=.85\columnwidth]{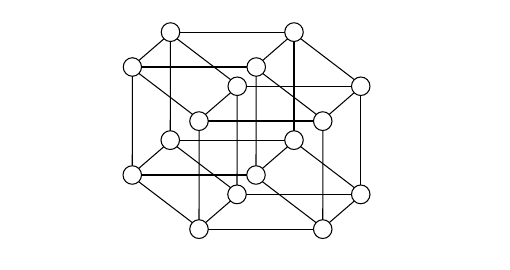}
    \vspace{-.25in}
    \caption{Hypercube Topology}
    \label{fig:hypercube}
    \vspace{-.15in}
\end{figure}

Besides just comparing structural properties of each topology, we demonstrate this experimentally in Section~\ref{topology-datamovement} to seed our study of the effects of connectivity on computational efficiency. \colorhl[yellow]{Lattice, Hex-Lattice, and Heavy-Hex topologies and various gates have been examined \textit{independently} to study the efficiency of routing and decomposition algorithms, respectively}~\cite{murali2019noiseadaptive}. \colorhl[yellow]{In contrast, we demonstrate the value in considering both gate and topology \textit{together} to benefit quantum workloads given the state-of-the-art transpilation---\textit{i.e.,} decomposition, placement, routing---algorithms.} 

\section{Motivation}
\label{section-motivation}
\label{sec:motivation}
To develop a NISQ quantum computer with improved fidelity and scalability there are several factors which lead to the need for collaborative design methodology. First, it is desirable to design a machine that has a target gate type in which it is efficient to map relevant quantum algorithms. Second, it is important to provide a rich and flexible topology to minimize the need for $\sw$ gates.
From the observations in this section we guide the design of our proposed quantum system described in Section~\ref{section:modularsnail}.

\renewcommand{\tabcolsep}{6pt}


\subsection{Normalizing Native Gate Sets}
\label{section:normalization}
\label{sec:gate-normalization}
We idealize different physically realizable gates and establish how, with factors such as qubit coherence removed,
we might determine whether a basis gate is more computationally useful than another. As previously noted, we treat 1Q gates as negligible.

Unlike classical computing, where the primary concerns are performance and energy consumption,  
the principal concern for quantum calculation is fidelity of the gates which perform computation. Moreover, infidelity in quantum computers can come from different sources. Some are only present during the gate operation, for example, driving the qubits to unwanted/error states and the imprecision/instability of the control electronics. Other sources of error are always present, for example, the loss of information from bits due to decoherence and energy loss. Common measures of gate fidelity, such as those experimentally determined by randomized  benchmarking~\cite{magesan2012characterizing}, combine the two together, further confusing the issue.  However, if one source of error dominates over the other then the strategy for circuit design must change.  Qubits which are idle retain their coherence,
and a good figure of merit is just the \textit{total number of gates} in the circuit~\cite{mccourt2022learning}. In contrast, if time is the dominant source of error for all qubits in the system, then \textit{circuit duration} is the best figure of merit, irrespective of the number of gates involved~\cite{sheldon2016characterizing, rol2017restless, somoroff2021millisecond}.  To address these two scenarios we produce throughout the remainder of the paper two parallel datasets: first, the total gate count, and second, the critical path gate count \textit{i.e.} total circuit duration, for a given circuit size, topology, algorithm, and basis.

\begin{observation}
We consider decomposition efficiency of the native basis gates realized by different modulators to predict their relative success. The best choice basis gate is $\syc$, $\cnot$, and $\sqisw$ for $\fsim$, $\CR$, and SNAIL modulators, respectively. Both $\cnot$ and $\sqisw$ require at most three instances to implement an arbitrary 2Q gate, whereas the best known analytical decomposition for $\syc$ requires exactly four\cite{crooks2020gates}.
In NISQ machines, data movement via $\sw$ gates can dominate many algorithms which requires three uses of either $\cnot$ and $\sqisw$. However, for a random distribution of 2Q gates, the $\sqisw$ requires only two uses far more often than the $\cnot$~\cite{huang2021towards}, providing a slight information theoretic advantage. 

\label{ob:basis-gate}
\end{observation}

\vspace{-.1in}
\subsection{Impact of Topology on Data Movement}
\label{topology-datamovement}

\begin{figure*}[h]
    \centering
    \includegraphics[width=\linewidth]{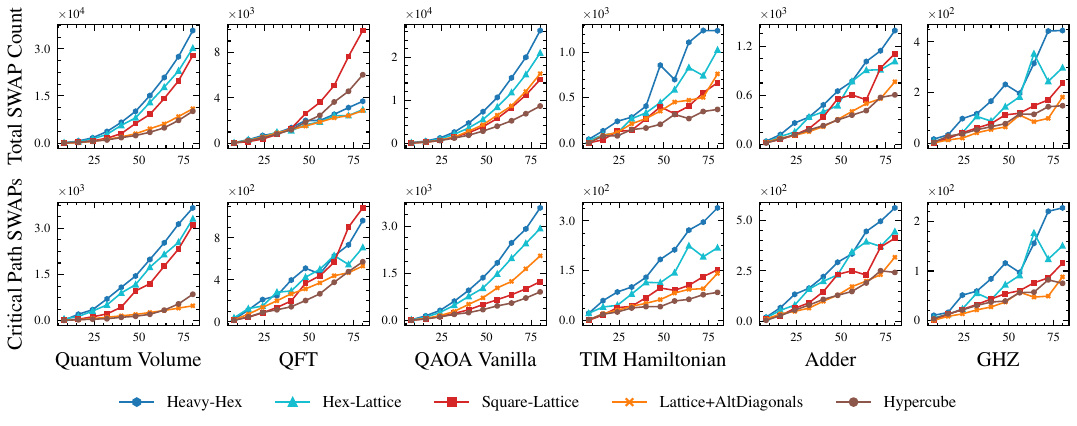}
    \caption{Total \textbf{(top)} and critical path \textbf{(bottom)} $\sw$ gates required for 80-qubit implementations on basic topologies. The count of induced $\sw$ gates is independent of the gate set and measures the efficiency of a topology subject to placement and routing transpilation passes.
    }
    \label{fig:motivation-topology-scaled}
    \vspace{-.1in}
\end{figure*}

We have transpiled several quantum benchmarks onto the set of 84-qubit topologies and observed the rqeuierd $\sw$ gates induced for data movement, independent of choice of basis gate, showin in Fig.~\ref{fig:motivation-topology-scaled}.Like before, Hex-Lattice and Heavy-Hex perform poorly in important benchmarks such as QV and QAOA. 
Unlike in the smaller problem sizes where the Square-Lattice is harder to distinguish from the richer topologies, it tends to follow trends of the hex configurations as the size scales. Only hypercube performs the best as size scales. Interestingly, while it does not dramatically reduce total $\sw$s for QFT, it scales comparatively better for critical path $\sw$s.  On average for an 80-qubit QAOA circuit, Heavy-Hex required $1.92\times$, $1.53\times$, and $2.83\times$ critical path $\sw$s more than Square-Lattice, Lattice+AltDiagonals, and Hypercube, respectively. This clearly demonstrates the need for richer topologies when scaling QC sizes. 

\begin{observation}
Unsurprisingly, topologies with higher connectivity generally scale better than sparse meshes.  However, topologies that prioritize reducing distance everywhere rather than dense neighborhoods of connectivity, i.e. avoiding bottlenecks of data movement, are more tolerant to scaling.
\label{ob:scaled-topologies}
\end{observation}

\vspace{-.15in}
\section{Quantum Co-Design with SNAILs}
\label{section:modularsnail}

Based on the observations in the prior section, there are several important factors to consider in the co-design a quantum architecture.  From observation~\ref{ob:basis-gate}, we should select a basis gate that minimizes the expected duration for decomposed 2Q gates.  From observation~\ref{ob:scaled-topologies}, we should construct a topology that efficiently scales in diameter while providing rich local connectivity. Collectively, our choice of basis gate should be designed collaboratively with a modulator that allows for increased qubit-qubit coupling neighborhoods. 

Next, we propose building a novel quantum architecture using the SNAIL quantum modulator.  Using the SNAIL allows natively implementing the $\sqisw[n]$ family and for a rich qubit couplings without frequency crowding. Additionally, the connectivity of SNAILs introduces the exploration of new topology configurations.

\subsection{SNAIL Parametric Modulator}
\colorhl[yellow]{Quantum mechanics describes how a state evolves in time by \textit{unitary} transformations (gates). The interactions of a system, specified by a Hamiltonian, yield a set of allowed energy eigenstates which determine the time-evolution unitary operator. SNAILs offer a way to control the Hamiltonian of superconducting circuit elements, such that the unitary transformations, \textit{i.e.,} quantum gates, \textit{are controlled}.}

The SNAIL is a flux-tunable device which can, for a certain applied flux, create a \textit{strong third-order} Hamiltonian term while canceling \textit{all $4^{th}$ and higher-order even terms}. The third-order term allows many discrete coupling frequencies while the fourth-order and higher even terms \textit{eliminated in the SNAIL} and \textit{found in the CR} and other modulators create cross-talk interactions which lead to \textit{frequency crowding}.
\begin{figure*}[h]
    \centering
    \includegraphics[width=.95\linewidth]{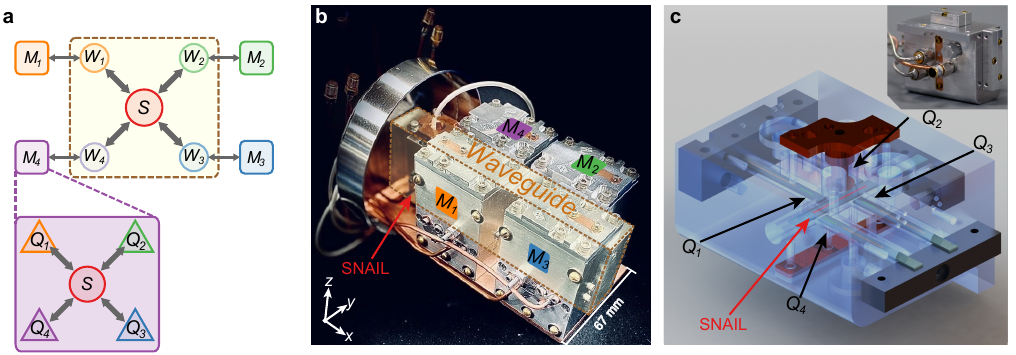}
    \caption{\textbf{Proposed modular quantum computer with quantum router and module} (a) Schematic of one quantum router coupled to four quantum modules. Each module has the same structure as module 4, forming a tree-like architecture. (b) Photograph of the SNAIL based quantum router and four simple modules (adapted from Ref.~\cite{lu2022quantum}). (c) Rendered representation and picture of the four qubit SNAIL-based quantum module.}
    \label{fig:hatlabRouterAndModule}
    \vspace{-.1in}
\end{figure*}
%

Thus the the Hamiltonian of SNAIL can be represented as:
\begin{equation}
    H_{SNAIL}/\hbar=\omega_{s}\hat{s}^\dagger\hat{s}+g_{3}(\hat{s}^\dagger+\hat{s})^3
\end{equation}
When coupled with other linear objects (\textit{e.g.,} harmonic oscillators) and non-linear objects (\textit{e.g.,} qubits), the total system inherits all possible three-body interaction terms from the SNAIL.
For instance, driving at the difference of two qubit resonant frequencies $\omega_{\textrm{drive}}=\omega_1-\omega_2$ creates the effective interaction: 
\vspace{-.1in}
\begin{equation}
H_{int}^{eff}= g_{12}^{eff} (\hat{a_1}^\dagger\hat{a_2} + \hat{a_1}\hat{a_2}^\dagger)
\end{equation}
\colorhl{The $\hat{a_1}^\dagger\hat{a_2}+\hat{a_1}\hat{a_2}^\dagger$ term creates an $\isw$ relationship between qubits 1 and 2 with a rotation intensity governed by $g$. If $g=\frac{\pi}{2n}\text{rad}$ the unitary/gate U =} \isfam \colorhl{and follows the transformation matrix:}
\vspace{-.05in}
\begin{equation}
    U(t) = e^{i H t / \hbar} =     \begin{bmatrix}
    1 & 0 & 0 & 0\\
    0 & \cos({gt}) & \mathit{i}\sin({gt}) & 0\\
    0 & \mathit{i}\sin({gt}) & \cos({gt}) & 0\\
    0 & 0 & 0 & 1
    \end{bmatrix}
\end{equation}

\colorhl[yellow]{Additionally, a stronger third-order term results in a higher coupling strength $g$, which is inversely proportional to time. In other words, it gives a stronger pump power with a faster rate of gate, reducing errors due to decoherence loss.}

SNAILs are based on the concept of parametric coupling.  
IBM's $\CR$ modulator uses fixed capacitive coupling and Google's $\fsim$ modulator uses tunable coupling.  
The parametric coupling idea of SNAIL modulators has long been used in parametric amplifiers~\cite{bergeal_analog_2010,sliwa2015reconfigurable,lecocq2017nonreciprocal}, and has recently been used to demonstrate qubit-qubit~\cite{noguchi2020fast}, cavity-cavity~\cite{gao2019entanglement}, and qubit-cavity~\cite{burkhart2020error,narla2016robust} gates.  
However, the SNAIL~\cite{frattini20173} has been designed to \textit{increase the frequency difference to several GHz} for distinguishing qubit-qubit coupling pairs, which increases their \textit{resilience against frequency crowding}.


In this system, which gate is produced from driving the SNAIL is controlled strictly by frequency selection; to create an addressable series of gates among many modes each must have a unique difference frequency not shared by another term in the Hamiltonian. 
Compared to the $\CR$ modulator, which requires a strong cross-Kerr term to operate, third-order parametric gates have much smaller cross-Kerrs 
due to static-$Z_1 Z_2$ interactions. Thus, SNAIL modultors \textit{allow operation of multiple gates in parallel} in the same neighborhood, or even create three- or more-mode ($\geq$3Q) gates by applying multiple, simultaneous drivers to the SNAIL.

Combined with low frequency crowding, these features of the SNAIL modulator allow flexible, parallel topologies with many to many superconducting qubit interactions, even across modules each with their own SNAILs.  Thus, some qubits can participate in multiple modules as topologies scale.


\vspace{-.05in}
\subsection{Example SNAIL Quantum Computer}



\begin{figure}[h]
    \vspace{-.1in}
    \centering
    \includegraphics[width=0.85\columnwidth]{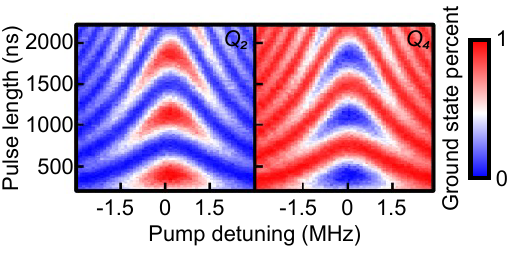}
    \caption{Parametrically driven exchange between two qubits, $Q_2$ and $Q_4$ of the quantum module}
    \label{fig:hatlabModuleSWAP}
    \vspace{-.2in}
\end{figure}

To demonstrate and characterize the capabilities of a superconducting quantum computer using SNAIL modulators, \colorhl[yellow]{in previous work}~\cite{zhou2021modular}, we constructed a quantum state router and a number of quantum modules with an overall architecture shown in Fig.~\ref{fig:hatlabRouterAndModule}a. In each module, four qubits are directly coupled to the same SNAIL using unique frequency modes allowing \isfam gates between each qubit pair.  

The router is made with a SNAIL chip placed inside a 3D superconducting waveguide. The SNAIL and the $TE_{10k}$ modes of the waveguide all couple to the SNAIL.  Thus all qubits in module $M_k$ can couple with waveguide mode $W_k$.  $W_k$ is coupled both to the SNAIL in $M_k$ as well as the SNAIL in the quantum state router and can form a gate with any element in either the module or quantum state router.  In our system $W_k$ is constructed as a cavity, which has slightly different properties than a qubit, but can still function to build \isfam gates.  Conceptually, $W_k$ can also be built as a qubit.  




Fig.~\ref{fig:hatlabRouterAndModule}b shows our preliminary physical implementation of the ``Tree,'' only with each module replaced by a simpler design to evaluate the performance of the router independently. In future experiments, an advanced module prototype depicted in Fig.~\ref{fig:hatlabRouterAndModule}c will be coupled to the router, which, together form the Tree architecture as depicted in Fig.~\ref{fig:hatlabRouterAndModule}a.
In each module, a SNAIL couples to all qubits for intra-module communication. Then, the modules are connected to the central SNAIL in the router through the piece marked as the waveguide.

Currently, the 4-port state router and 4-qubit module sub-systems have been physically realized in two separate experiments.  Preliminary results of the router~\cite{lu2022quantum} have demonstrated all-to-all exchange interactions among four modules. For 4-qubit module experiment, representative data of one qubit-qubit exchange is shown in Fig.~\ref{fig:hatlabModuleSWAP}, which shows an excitation swapping between qubits when the SNAIL is pumped at different durations and detunings. \colorhl[yellow]{This figure depicts how gate operations are continuous in time, as we see each qubit oppositely alternating states along the y-axis duration of the gate. Also, it depicts a sense of the fidelity of the operation, as the darker hues represent a purer distribution of measured state outcomes.} We have demonstrated that the router is capable of performing $\isw$ family gates and can create entanglement between arbitrary qubit pairs from different modules, with an average inter-module gate fidelity of $\sim 97\%$. We have also shown that by keeping the $W_k$ device empty, it is possible to build $\sw$ gates using a single $\isw$.  

\colorhl[yellow]{The primary fidelity limit in this device is the ratio of gate time to qubit coherence time} \cite{zhou2021modular}. \colorhl[yellow]{Qubit lifetimes depend on their internal loss which comes from many sources such as loss from the metal package or drive ports as well as through coupling to the SNAIL. One source of decoherence is due to flux-noise, which can detune the SNAIL and cause qubits to dephase.  Addressing these challenges requires engineering effort to move from devices created in the lab to industry research products on their way to commercialization.  With a similar amount of effort SNAILs can reach pulse speeds and decoherence of other modulators that have received this engineering investment.  Thus, to evaluate that core potential of these approaches without overemphasizing the engineering effort a particular modulator or implementation has received, we normalize the experimental evaluation to duration in terms of pulses, gate count, etc.  This allows a comparison of the potential to help guide the validity of further investment in modulators like the SNAIL the potential for high-fidelity qubit-couplings, and modular structures to advance the state-of-the-art of NISQ QCs.} 


\subsection{SNAIL-based Topologies}
\label{sec:snail-topologies}
\begin{figure}[h]
\vspace{-.1in}
    \centering
    \subfloat[Modular 4-ary Tree]{
    \includegraphics[width=.48\columnwidth]{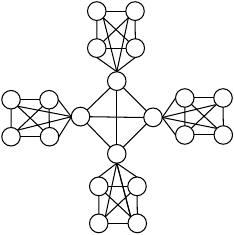}
    \label{fig:mod-tree}
    }
    \subfloat[Round Robin 4-ary Tree]{
    \includegraphics[width=.48\columnwidth]{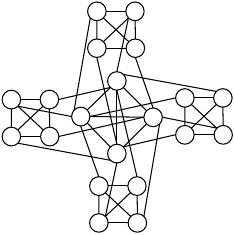}
    \label{fig:rr-tree}
    }
    \caption{Two-level 4-ary (20-qubit) tree topologies from SNAILs.}
    \label{fig:tree-topologies}
    \vspace{-.1in}
\end{figure}

Based on the demonstrated prototype quantum computer from Fig.~\ref{fig:hatlabRouterAndModule}, \colorhl[yellow]{in our new research,} we have extrapolated feasible topologies that can be realized using the SNAIL modulator.  The basic design is the 20-qubit Tree topology, seen in Fig.~\ref{fig:mod-tree}.  The major difference between this topology and the demonstrated prototype system in Fig.~\ref{fig:hatlabRouterAndModule}a is that central nodes, which are analogs to the $W_k$ elements, are considered to be qubits rather than cavities, as noted previously.  However, the standard Tree design contains bottlenecks in the router qubits.  Thus, we explore a theoretical alternative design where each qubit in a neighborhood $k$ is connected to each $W_0\ldots W_3$ in ``round robin'' fashion.   The goal is to eliminate the bottleneck of the $W_k$ qubits, to allow data to move more easily between neighborhoods in parallel.  This design, called a ``Round Robin Tree'' (Tree-RR) decreases the maximum distance between all pairs.  


\begin{figure}
    \centering
    \includegraphics[width=.8\columnwidth]{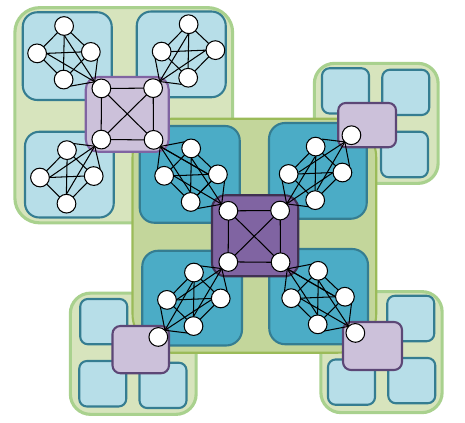}
    \caption{\colorhl[yellow]{Four-level 4-ary tree (84-qubit) tree topology}}
    \label{fig:tree-84}
\end{figure}

We append additional levels, where each module is connected to the following level's router, to create 84-qubit versions of both the Tree, depited in Fig.~\ref{fig:tree-84}and Tree-RR topologies. For the 84-qubit Tree-RR, each module couples to a different second-level router qubit, and each second-level router qubit is coupled to a different first-level router qubit as shown in Fig.~\ref{fig:rr-tree}.

As we explored the potential of the hypercube topology in Section~\ref{sec:motivation}, we explore topologies enabled by the SNAIL with similar properties.  
\colorhl[yellow]{To avoid frequency crowding, a SNAIL can typically interact among as many as six qubits. For our modular design, a module will contain one SNAIL and up to four qubits. SNAILs interact with all qubits in their module. Then to realize larger topologies, qubits link between modules and are coupled with the neighboring module's SNAIL and qubits.
We see this in the Tree, with each module containing four qubits and one SNAIL connected to a shared qubit to cross module boundaries.
}

\begin{figure}[h]
    \centering
    \subfloat[Corral$_{1,1}$ SNAIL-Qubit Coupling Diagram]{
    \includegraphics[width=.45\columnwidth]{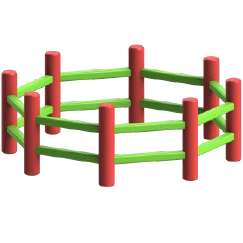}
    \label{fig:corral-a}
    }
    \hfill
    \subfloat[Corral$_{1,1}$ Topology]{
    \includegraphics[width=.45\linewidth]{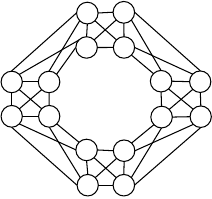}
    \label{fig:corral-b}
    }\\
    \subfloat[Corral$_{1,2}$ Coupling Diagram]{
    \includegraphics[width=.45\columnwidth]{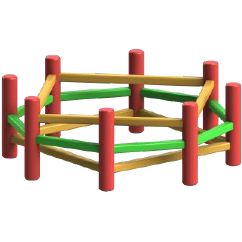}
    \label{fig:corral-c}
    }
    \hfill
    \subfloat[Corral$_{1,2}$ Topology]{
    \includegraphics[width=.45\linewidth]{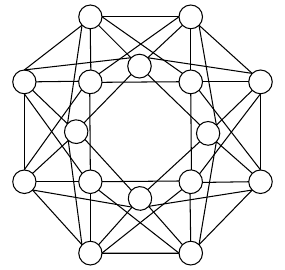}
    \label{fig:corral-d}
    }
    \caption{Topologies from SNAIL Corral architectures}
    \label{fig:corral-topologies}
    \vspace{-.1in}
\end{figure}

Using this constraint, it was desirable to build a topology that maintained the low-diameter property of hypercubes without the required connectivity dimension scaling.  Thus we proposed a Corral structure inspired by the 4D hypercube and explored an analog of Tree and Tree-RR for these Corrals. In Fig.~\ref{fig:corral-topologies}, red vertical cylinders represent SNAILs and green or yellow horizontal bars are the qubits coupled between them. 
We call them Corrals due to their resemblance to fence-posts. By building a octagonal ring of modules, each SNAIL with 2 levels of qubits, is defined by a pattern of fence-post connections. Whereas Fig.~\ref{fig:corral-a} is the easiest to physically realize, with each qubit coupled to the nearest adjacent SNAIL, denoted Corral$_{1,1}$, we might also realize differing strides, which is reminiscent of the hypercube. The Corral$_{1,1}$, shown in Fig.~\ref{fig:corral-b} creates groups of 4 qubit all-to-all coupling, whereas Corral$_{1,2}$ shown in Fig.~\ref{fig:corral-c}, connects its second fence to the second-nearest neighbor, decreasing the average distance between all pairs of qubits. The resulting topology is shown in Fig.~\ref{fig:corral-d}. 

Despite initially appearing similar to a ring, we note that the Corral topologies actually exhibit an coupling in neighbors (5 or 6 qubits) analogous to a 4-D hypercube of the same size (4 qubits). 

The module from \ref{fig:hatlabRouterAndModule}c is realized in a 3-dimensional structure.  If we define the Corrals as modules with two qubits and a SNAIL, we can also realize these in a 3-dimensional structure resembling Fig.~\ref{fig:corral-a} and \ref{fig:corral-c}.  For instance, presume each SNAIL (the fence-post) and the two qubits to the right are part of each module.  Then the SNAIL from the the right module can also control this module's qubits, and the SNAIL from this module can also govern the left module's qubits.  Alternatively, we can create heterogeneous modules where one module ($mod_1$) contains a SNAIL and four qubits, and another module ($mod_2$) contains only a SNAIL that forms the boundary between two $mod_1$'s.  Thus, Corrals can be scaled by adding more posts (modules) in the ring.  However, other approaches could be to combine Corrals with Tree-like modules or to layout Corrals in a lattice pattern.

One benefit of SNAILs is that they isolate the set of distinguishable qubit frequencies in a scope, which is not limited to within a module.  Thus, inter-module frequencies have the same fidelity as intra-module frequencies.  In the next section we describe our methodology to compare SNAIL-based NISQ QCs with systems using CR (IBM) and FSIM (Google) modulators.


\section{Experimental Setup}
To compare the SNAIL-based superconducting quantum computer with machines from IBM and Google we use the metrics for normalization as described in Section~\ref{sec:gate-normalization}.  For simplicity, we presume all gates have uniform fidelity. We explore two sizes of machine.  One of the scope for which we have a hardware prototype with 16-20 qubits.  These designs are summarized in Table~\ref{tab:small-topologies}.  The table includes the number of nodes (qubits), the diameter of the topology (Dia.), the average distance between any two qubits (Avg$_\text{D}$), 
and the average connectivity of a qubit (Avg$_\text{C}$).  Tree, Tree-RR, Corral$_{1,1}$, and Corral$_{1,2}$, are all topologies realizable by the SNAIL. Square-Lattice is included as a baseline and hypercube is included for comparison against the corrals.

\begin{table}[h]
\centering
    \caption{Topologies and Connectivities}   
        \begin{tabular}{c|c|c|c|c}
        \hline\hline
             & \textbf{Qubits} & \textbf{Dia.} & \textbf{Avg$_\text{D}$} & \textbf{Avg$_\text{C}$}\\
             \hline
             Heavy-Hex & 20 & 8.0 & 3.77 & 2.1\\
             Hex-Lattice & 20 & 7.0 & 3.37 & 2.45 \\
             Square-Lattice & 16 & 6.0 & 2.5 & 3.0\\
             Tree & 20 & 3.0 & 2.15 & 4.6  \\
             Tree-RR & 20 & 3.0 & 2.03 & 4.6  \\
             Corral$_{1,1}$ & 16 & 4.0 & 2.06 & 5.0  \\
             Corral$_{1,2}$ & 16 & 2.0 & 1.5 & 6.0  \\
             Hypercube & 16 & 4.0 & 2.0 & 4.0\\\hline\hline
        \end{tabular}
    \label{tab:small-topologies}
\end{table}
\begin{table}[h]
\centering
    \caption{Scaled Topologies and Connectivities}
        \begin{tabular}{c|c|c|c|c}
             \hline\hline
             & \textbf{Qubits} & \textbf{Dia.} & \textbf{Avg$_\text{D}$} & \textbf{Avg$_\text{C}$}\\
             \hline
             Heavy-Hex & 84 & 21.0 & 8.47 & 2.26\\
             Hex-Lattice & 84 & 17.0 & 6.95 & 2.71\\
             Square-Lattice & 84 & 17.0 & 6.26 & 3.55\\
             Lattice+AltDiag & 84 & 11.0 & 4.62 & 5.12\\
             Tree & 84 & 5.0 & 3.91 & 4.71  \\
             Tree-RR & 84 & 5.0 & 3.65 & 4.71 \\
             Hypercube & 84 & 7.0 & 3.32 & 6.0\\\hline\hline
        \end{tabular}
\label{tab:scaled-topologies}
\end{table}

We scale the size of the machine to a system with 84 qubits as described in Table~\ref{tab:scaled-topologies}.  The Tree and Tree-RR are expanded with a third level router as described in Section~\ref{sec:snail-topologies}.  \colorhl[yellow]{We retain the hypercube as a Corral-like design to see the potential of building larger scaled Corrals in addition to scaled} Heavy-Hex, Hex-Lattice, Square-Lattice, and Lattice+AltDiag as comparison points.  While it is relatively straightforward to remove nodes from the lattice patterns to match 84 nodes, for the Hypercube we must a node from each dimension to reduce the size while maintaining the regular structure.  


\colorhl[yellow]{To generate circuits for these QCs, we extended the Qiskit Terra 0.20.0 transpiler.  We provided new backends that support analytical $\sqisw$  and $\syc$ decompositions using Cartan's KAK method. We also extended the transpiler to include the our proposed Tree, Tree-RR, Corrals, Hypercube, and Lattice+AltDiagonals topologies. For design-space exploration we ensure that each basis gate can be assigned to each topology. Note Qiskit already includes an Carton's KAK $\cx$ decomposition backend, as well as Square-, Hex-Lattice, and Heavy-Hex topologies.}
We use Qiskit's DenseLayout for initial qubit mapping and StochasticSwap for routing $\sw$s.

We tested the machine configurations with workloads that include widely used quantum circuits that can be scaled to different problem sizes.  
Our parameterized circuits are QuantumVolume, QFT, and CDKMRippleCarryAdder from Qiskit and QAOAVanillaProxy, HamiltonianSimulation, and GHZ from Supermarq~\cite{tomesh2022supermarq}. 
\colorhl[yellow]{We select these circuit benchmarks over other popular algorithms, such as VQE, because they can be parameterized as a function of qubit size and be generated automatically, while VQE and other similar benchmarks would require hand-coded designs for all problem sizes.} 

\begin{figure}
    \centering
    \includegraphics[width=\columnwidth]{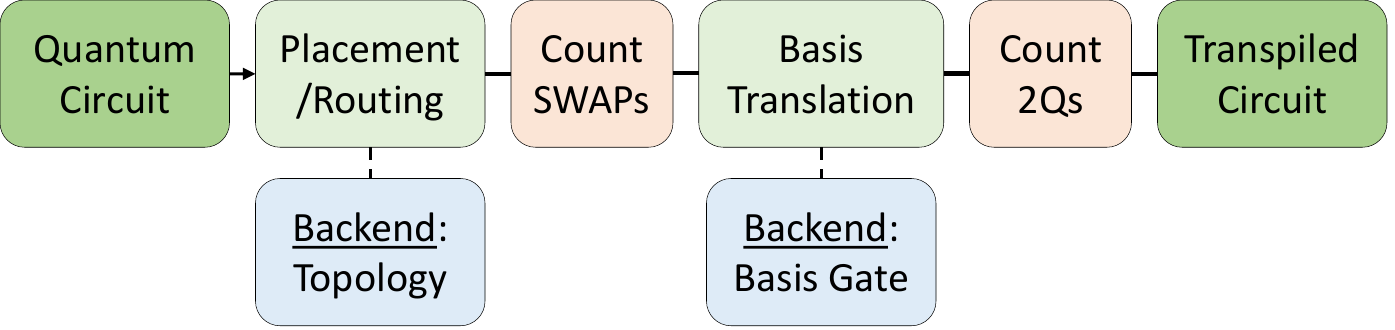}
    \caption{\colorhl[yellow]{Flow of circuit transpilation and data collection}}
    \label{fig:eval-method-loop}
    \vspace{-.1in}
\end{figure}

\colorhl[yellow]{During transpilation }(Fig.~\ref{fig:eval-method-loop}) \colorhl[yellow]{we collect 4 sets of data over each backend, for each circuit of incremental size. We use Qiskit's functionality
to count total gates and critical path gates. After the routing pass, we count the total induced $\sw$ and critical path $\sw$ gates. After the final basis translation pass, we count the total 2Q gates and critical path 2Q gates.} \colorhl[yellow]{Note, we use 2Q basis gate count and its associated pulse duration as a surrogate for determining overall reliability as described in Section~}\ref{section:normalization}. These experimental settings were used to generate the charts shown in Section~\ref{sec:motivation} as well as the experimental results that we describe next.

\begin{figure*}[h!]
    \centering
    \includegraphics[width=\linewidth]{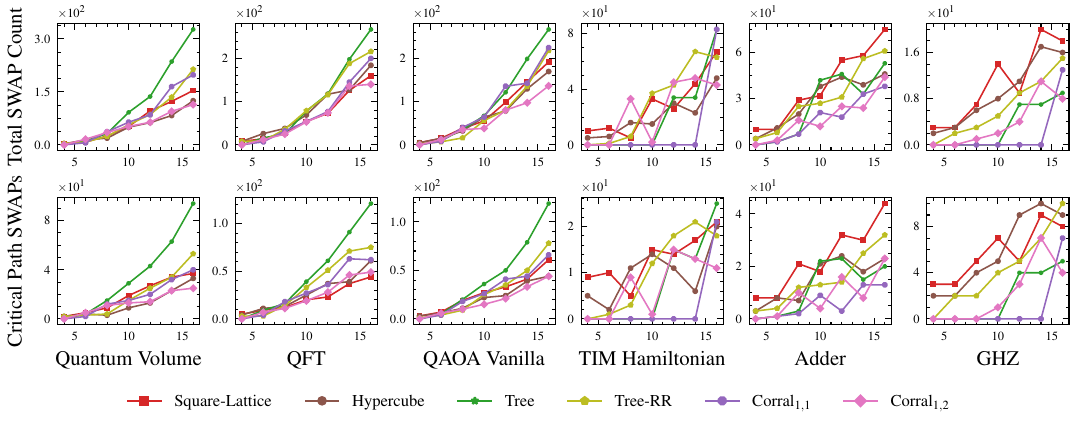}
    \caption{Total \textbf{(top)} and critical path \textbf{(bottom)} f$\sw$ gates required for 16-qubit implementations of proposed SNAIL topologies.}
    \label{fig:results-swaps-corral}
    \includegraphics[width=\linewidth]{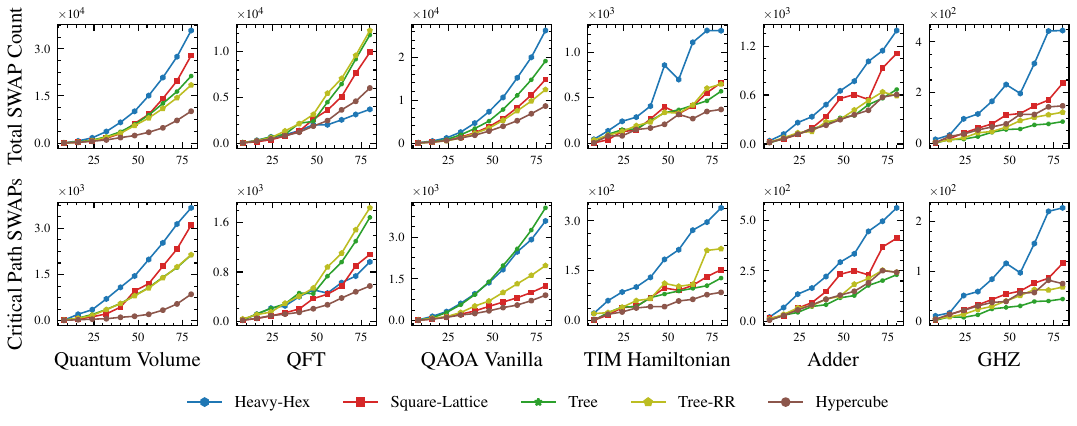}
    \caption{Total \textbf{(top)} and critical path \textbf{(bottom)} $\sw$ gates required for 84-qubit implementations comparing proposed SNAIL topologies against common topology baseline.}
    \label{fig:results-data-treeswaps}  
    \vspace{-.1in}
\end{figure*}

\vspace{-.1in}
\section{Results}
To evaluate the potential of the SNAIL QCs we explore the impact of the newly possible topologies similarly to Sec.~\ref{sec:motivation}, in a gate agnostic fashion, then combine the topology and gate impacts to represent the co-design advantage.  Finally, we explore the potential of other basis gates in the \isfam family by investigating fidelity of $n > 2$ decompositions.


\subsection{Evaluation of SNAIL-enabled topologies}
We evaluate the proposed SNAIL-coupling corral topologies against the previously discussed topologies, shown in Fig.~\ref{fig:results-swaps-corral}. The scaling of total and critical $\sw$ gates moderately obeys the expected ordering, as when average connectivity goes up and average distance goes down, less $\sw$ gates are required. Despite the smaller circuits not quite converging to steady trends, excitingly, the corral topologies are still unambiguously the best performers. Noticeably, the transpiler manages to find an initial mapping that often requires zero $\sw$ gates for Corral$_{1,1}$, an indicator of its rich connectivity. 

To extrapolate to larger topologes, we revisit the topologies from Section~\ref{topology-datamovement}, now including the SNAIL topologies Tree and Tree-RR, shown in Fig.~\ref{fig:results-data-treeswaps}. Hex-Lattice and Lattice+Diag are are not shown for readability and redundancy. This gives us a comparison relating to the previous set of benchmarks. Once again, the constant properties of the topologies appear to generally coincide with performance. In fact, for an 80-qubit QV circuit, we compute from Heavy-Hex to Tree a 54.3\% decrease in total $\sw$ gates or a 79.8\% decrease in critical path $\sw$ gates. However, the Tree designs do not quite match the performance of the hypercube, as from Tree to hypercube experiences an additional 42.5\% decrease in total $\sw$ or 54.3\% decrease of critical path $\sw$ gates.

\subsection{Evaluation of Collaborative Design}
\begin{figure*}[h!]
    \centering
    \includegraphics[width=\linewidth]{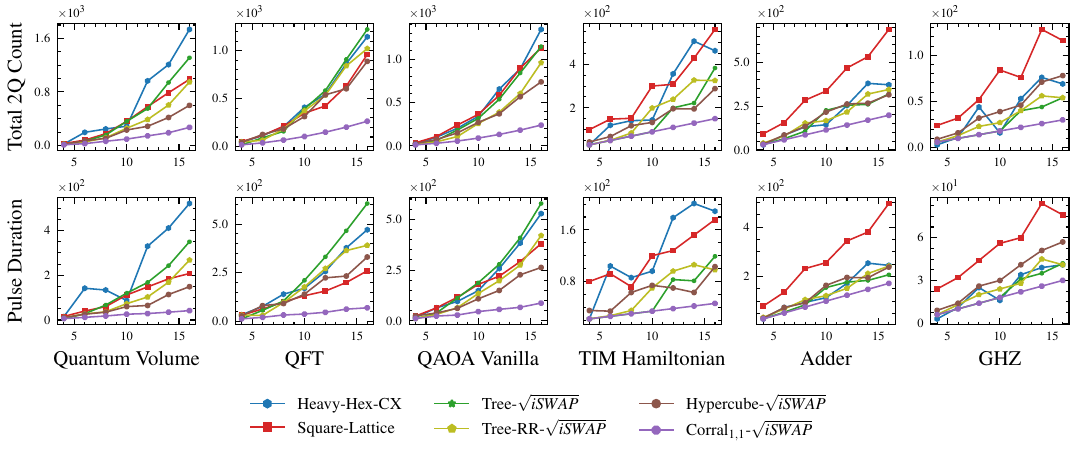}
    \caption{\colorhl{Total \textbf{(top)} and critical path \textbf{(bottom)} 2Q gate counts, decomposed into the respective native basis sets, required for 16-qubit implementations comparing proposed SNAIL topologies against common topology baseline.}}
    \label{fig:results-smallpulseduration}
    \centering
    \includegraphics[width=\linewidth]{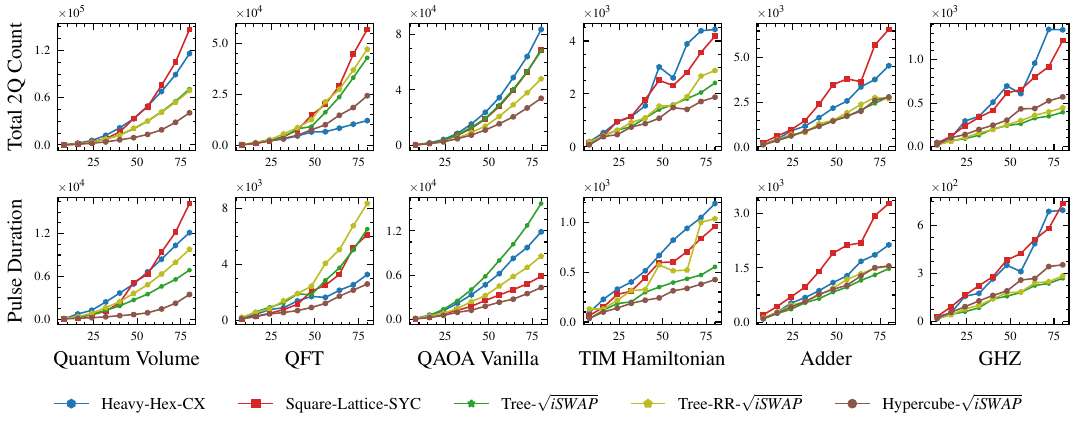}
    \caption{Total \textbf{(top)} and critical path \textbf{(bottom)} 2Q gate counts, decomposed into the respective native basis sets, required for 84-qubit implementations comparing proposed SNAIL topologies against common topology baseline.}
    \label{fig:results-pulseduration}
    \vspace{-.1in}
\end{figure*}

Next, we continue the decomposition into each topology's basis gate, in order to count the total number of 2Q gates and the number of 2Q gates in the critical path, \textit{i.e.} the pulse duration of the circuit, shown in Fig.~\ref{fig:results-smallpulseduration} and Fig.~\ref{fig:results-pulseduration}. As noted before, each basis gate is approximately the same with some preference going to $\sqisw$ over $\cnot$, and $\cnot$ over $\syc$, effectively adding a scaling factor enough to bring the $\syc$ gate on Square-Lattice above the $\CR$ gate on Heavy-Hex.


This data also exhibits important principles of parallelism on the topologies, meaning when a curve flattens from total gate count to duration, that means more gates are not contributing to the duration and therefore are in parallel time steps. As an example, the Tree on the QV benchmark flattens it total 2Q gate count onto the duration plot, suggesting a comparatively high degree of gate parallelism. Of particular note, the co-design method shows that the Corral topology combined with the $\sqisw$ enabled with the SNAIL modulator consistently outperforms all other designs for all benchmarks.



When scaling to larger topologies, Fig.~\ref{fig:results-pulseduration} we see interesting trends. Heavy-Hex scales the worst for QV, and the best for QFT, meanwhile, Tree-RR scales the worst for QFT and the best for GHZ. This solidifies the variability of different applications on a topology and should temper excitement about a singular benchmark (even QFT) succeeding on a topology without evidence that other workloads also have positive trends. Interestingly, the hypercube is generally among the best for all benchmarks which lends support to further development of its inspired topologies like Corral. 
Note, the transpilation placement and routing heuristics appear noisy with some problem sizes more naturally embedding $G'$ into $G$s than others. This may explain why gate counts are not always monotonic on the same topologies and could be a source of inefficient use of more rich topologies.

Thus far we have investigated a set of qubit coupling modulators and both the native gate basis and topology architectures they produce. 
Most importantly, that Tree topologies make a large improvement in connectivity without needing to sacrifice gate fidelity. We identified decreasing connectivity as a harmful trend in topology design and studied an aspirational hypercube topology for benchmarking richer realizable toplogies to address connectivity, such as Tree-RR and a Corral. 
\subsection{Pulse Duration Sensitivity Study}
\label{section:approxdecomp}
As previously discussed, using the SNAIL to realize gate couplings yields a \isfam family of gates. $\sqisw$ has been studied as a naturally good candidate for forming a basis, which already decreases the basis gate pulse duration by $\frac{1}{2}$.  However, exploring even smaller fractions of $\isw$ for decomposition to decrease duration and decoherence time have not been studied. While $\sqisw$ is the smallest fraction that is a ``perfect entangler''~\cite{huang2021towards}, decompositions to $\sqisw[n]$ where $n>2$ may result in high-fidelity decompositions. However, no analytical decomposition to these gates has been discovered, thus we use an approximate decomposition engine to explore this possibility.

We reproduce a version of NuOp \cite{nuop} to build template circuits, which interleave the desired \isfam gate with 1Q gates (Eq.~\ref{synthesis:template}) similar to the exact decomposition method in Section~\ref{section:decomp}. However, because the decomposition is approximate, this introduces another form of error beyond decoherence, the error from the decomposition approximation.  Thus, \colorhl{the similarity between unitaries, used as the decomposition fidelity is defined using the Hilbert-Schmidt inner product between the template and target from Eq.}~\ref{eq:hilbertschmidt}.

\begin{equation}\label{synthesis:template}
\resizebox{\columnwidth}{!}{
\Qcircuit @C=1.0em @R=0.4em @!R {
	 	& \multigate{2}{\mathrm{U}} & \qw & & & \gate{\mathrm{U}} & \multigate{2}{\mathrm{\sqrt[n]{\texttt{iSWAP}}}} & \dots & & \multigate{2}{\mathrm{\sqrt[n]{\texttt{iSWAP}}}} & \gate{\mathrm{U}} & \qw
	 	\\
	 	& & & \push{\rule{0em}{0em}\!\!\!\!\approx\!\!\!\!\rule{0em}{0em}}
	 	\\
	 	& \ghost{\mathrm{U}} & \qw & & & \gate{\mathrm{U}} & \ghost{\mathrm{\sqrt[n]{\texttt{iSWAP}}}} & \dots & & \ghost{\mathrm{\sqrt[n]{\texttt{iSWAP}}}} & \gate{\mathrm{U}} &\qw\\
}}
\end{equation}
\vspace{-.15in}
\begin{equation}
    \label{eq:hilbertschmidt}
    F_d(U_d, U_t) = \frac{\text{Tr}(U_d^{\dagger}U_t)}{\text{dim}(U_d)}
\end{equation}

However, given our goal is to improve fidelity by reducing decoherence, we also approximate that decoherence scales linearly over time, as shown in Eq.~\ref{eq:infidelity}. To illustrate, consider an $\isw$ some base duration that increases decoherence such that fidelity reduces to $90\%$.  A gate with half the duration has approximately half the decoherence, hence infidelity is reduced from $10\%$, to $5\%$, yielding a $95\%$ fidelity from the duration of a $\sqisw$ gate.
\begin{equation}
    \label{eq:infidelity}
    F_b(\sqrt[n]{\text{\texttt{iSWAP}}}) = 1 - \frac{1 - F_b(\isw)}{n}
\end{equation}


%

As a result, the best total fidelity of the unitary decomposition is the product of the total sum delay of $k$ applications of the basis gate (max$_k$) and the infidelity of the approximation of each of the $k$ basis gates in the decomposition described in Eq.~\ref{eq:total_fidelity}. We ignore the delay of the 1Q gates as in previous decomposition.  To study this we generated circuits for the Haar distribution of 2Q unitaries and to find the best total fidelity, we iterate the template size $k$ based on Eq.~\ref{eq:total_fidelity}.
\vspace{-.08in}
\begin{equation}
    \label{eq:total_fidelity}
    F_t = \max_k F_d^{(k)} (F_b)^k
\end{equation}


\begin{figure}[h]
    \centering
    \includegraphics[width=.9\columnwidth]{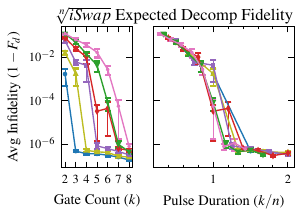}
    \vspace{-.1in}
    \includegraphics[width=.9\columnwidth]{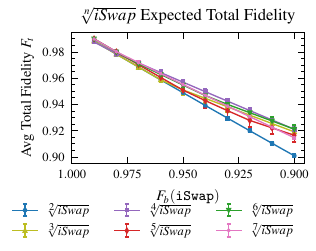}
    \caption{\colorhl[yellow]{Fidelity comparisons for Haar-random 2Q unitaries on a set of} \isfam \colorhl{basis gates (N=50).}}
    \label{fig:scaling}
\end{figure}

As evidenced by Fig.~\ref{fig:scaling}(top left), smaller fractional \isfam need more repetitions (larger $k$) to reach near-exact decompositions, visible by reaching $<10^{-6}$ fidelity with higher values of $k$. However, the savings pulse duration subceeds the larger number of gates, reducing the total pulse duration. For example, given high-fidelity decomposition for $\sqisw$ in $k=3$ and $\sqisw[3]$ in $k=4$, the total duration is reduced from 1.5 to 1.33. This is verified again in Fig.~\ref{fig:scaling}(top right), where as $n$ grows, the total pulse duration decreases.
Fig.~\ref{fig:scaling}(bottom) shows the total fidelity of decoherence and approximate decomposition with decoherence due to $\isw$ pulse length on the x-axis and total fidelity on the y-axis.  We find that for Haar-randomly sampled 2Q unitaries and for a $99\%$ fidelity $\isw$ basis, for $k = \{3,4,5\}$ $\sqisw[k]$ decreases infidelity by $14\%$, $25\%$, and $11\%$. 
This evidence continues to support the SNAIL modulator for its realization of a powerful basis set, with the ability to modify duration of the continuous operator to maximize gate fidelities.  


\section{Conclusion}
\vspace{-.05in}
In this work, we demonstrate the data movement overheads and penalties from lattice-based NISQ machines on a range of algorithms.  The SNAIL-based modulator and modular architectures provide significant improvements over 2D lattices, particularly for smaller node sizes, whereas a directly scalable 4-ary Tree structure shows mixed performance over the benchmark workloads for larger node sizes.  The central node can create bottlenecks for QFT and QAOA which require frequent long-distance connections.  However, the Tree performs very well for the Quantum Volume and GHZ-creation tasks due to their rich local connections.  
We also observe that a hypercube structure, which has rich local connections and low diameter, is superior to both lattices and the Tree and its variants, with exception of the highly ordered GHZ-state creation. We found that on an average of Quantum Volume circuits ranging from 16 to 80 qubits, a hypercube topology induces $2.57\times$ less total $\sw$ gates and $5.63\times$ less critical path $\sw$ gates compared to Heavy-Hex. We have explored hypercube inspired `Corral' structures which are both feasible given the demonstrated capabilities of SNAIL modulators for 16 nodes and provide superior computational performance particularly when coupled with $\sqisw$ (see Fig.~\ref{fig:results-smallpulseduration}). 

All of our connectivity designs are physically realizable with our demonstrated SNAIL modulators, and represent excellent targets for ours and others' next-generation quantum computers.  Our results point to the need for both dense connectivity and a mix of short- and long-range links in future NISQ machines. 
Finally, the strong performance of $\sqisw$, which is native to the SNAIL modulator, inspired us to explore whether smaller fractions of \isfam can yield a superior approximate implementations. We found that for 2Q unitaries and for a $99\%$ fidelity $\isw$ basis, $\sqisw[4]$ decreases infidelity on average by $25\%$ compared to $\sqisw$, leading to further co-design advancements.  

In future work, determination of analytical decompositions for $\sqisw[n]$ where $n>2$ can benefit decoherence time without introducing approximate decomposition error.  Additionally, exploration of heterogeneous basis gates to further reduce pulse time is an important direction.  Finally, exploring methods to scale Corral or develop new SNAIL realizable topologies to compete with aspiration hypercube topologies for larger qubit numbers is an important next step.

  

\section*{ACKNOWLEDGMENTS}
\vspace{-.1in}
This work is partially supported by the Laboratory of Physical Sciences and NSF Award CNS-1822085.  MX, MJH, and PL are partially supported by the U.S. Department of Energy, Office of Science, National Quantum Information Science Research Centers Co-Design Center for Quantum Advantage under contract DE-SC0012704.  


\bibliographystyle{IEEEtran}
\bibliography{refs}
\end{document}